\documentclass{sig-alternate}
\usepackage{textcomp}
\usepackage{multirow}
\usepackage{amssymb}
\usepackage{slashbox}
\usepackage{algorithm}
\usepackage{algorithmic}
\usepackage[normaltitle,normalsections,normalmargins,normalbib,normalbibnotes]{savetrees}

\newtheorem{theorem}{Theorem}
\newtheorem{definition}{Definition}

\newtheorem{lemma}{Lemma}

\begin{document}
%
\conferenceinfo{WOSS 2012}{Istanbul, Turkey}

\title{On Graph Deltas for Historical Queries}

\numberofauthors{3} 
%
\author{
%
\alignauthor
Georgia Koloniari\\
       \affaddr{University of Ioannina, Greece}\\
       \email{kgeorgia@cs.uoi.gr}
\alignauthor
Dimitris Souravlias\\
       \affaddr{University of Ioannina, Greece}\\
       \email{dsouravl@cs.uoi.gr}
\alignauthor
Evaggelia Pitoura\\
       \affaddr{University of Ioannina, Greece}\\
       \email{pitoura@cs.uoi.gr}
}

\maketitle
\begin{abstract}
In this paper, we address the problem of evaluating historical queries on graphs. To this end, we investigate the use of graph deltas, i.e., a log of time-annotated graph operations. Our storage model maintains the current graph snapshot and the delta. We reconstruct past snapshots by applying appropriate parts of the graph delta 
on the current snapshot. Query evaluation proceeds on the reconstructed snapshots but we also propose algorithms based mostly on deltas for efficiency. 
We introduce various techniques for improving performance, 
including materializing intermediate snapshots, partial
reconstruction and indexing deltas.  
\end{abstract}




\section{Introduction}
In recent years, there has been increased interest in graph structures representing 
real-world networks such as social networks, 
citation and hyperlink networks as well as biology and computer networks. 
In this paper, we focus on social network graphs.
Such graphs are characterized by large-scale, since the number of participating nodes 
reaches millions. Social graphs are also highly dynamic, 
since the corresponding social networks constantly evolve through time.

An interesting problem in this setting is supporting \textit{historical queries}. 
By historical queries, we refer to queries that involve the state of the graph 
at any time interval in the past. For instance, consider queries about 
the popularity (e.g., number of friends) of a user
at some specific time in the past, about how this popularity changed over time as well as
queries about the diameter of a network over a time period.                          
Historical queries are important when studying graphs that change through time 
for various applications such
as version maintenance and monitoring and analyzing the evolution of the graph. 

However, most recent research mainly addresses 
the problems introduced by the large-scale of social graphs \cite{KTSLF11, KTF11, MABDHLC10} 
and ignores the temporal aspects by focusing on queries that involve only the current graph snapshot. 
In this paper, we introduce a framework for supporting historical queries that 
involve one or more graph snapshots. To this end, we propose a general model for 
incorporating information about how a graph changes through time 
based on graph deltas. A \textit{graph delta} is a log of time-annotated graph update operations 
such as the addition and removal of nodes and edges. 
At each time instant, we store a current snapshot of the graph plus the graph delta 
that records the changes that have occurred. 
By applying the graph delta on the snapshot, any past snapshot can be reconstructed.
We also discuss materializing intermediate snapshots to improve performance.

The evaluation of historical queries is based on a two-phase plan 
that first reconstructs the snapshot 
or snapshots that are required to evaluate the query.  
We address the high cost of snapshot reconstruction by proposing query plans that rely only or mostly on the delta 
when this is possible. Furthermore, for node-centric queries, i.e., queries that access only parts of the graph, we introduce partial snapshot reconstruction that constructs only the subgraph required to evaluate the query. Finally, we show how 
building indexes on deltas can further improve efficiency.

The rest of this paper is organized as follows. 
Section 2 introduces our model for storing time-evolving graphs. 
Section 3 presents a classification 
of historical queries and query plans for their evaluation. 
Section 4 reports experimental results. Section 5 summarizes related work, while Section 6 concludes the paper.

\section{Model}
We model a social network as an undirected graph, $G= (V,E)$. Each graph node $v_i \in V$ corresponds to a user $u_i$ of the social network. Edges $(v_i,v_j) \in E$ capture social relationships (i.e., friendship) between users $u_i$ and $u_j$ that correspond to nodes $v_i$ and $v_j$ $\in V$ respectively.

Note that our model supports symmetrical social relationships between users, 
such as friendship in Facebook. If we consider asymmetrical relationships such as the ones in Twitter, then the graph representing the network is directed as the edges capturing the ``follower'' and ``following'' relationships in the network are also directed, i.e.,  
$u_i$ can ``follow'' $u_j$, while $u_j$ does not ``follow'' $u_i$. In the rest of the paper, we focus on undirected graphs but our algorithms can be easily adopted to account for directed graphs.
 
Our model for capturing the evolution of the social network through time is based on the use of \textit{graph snapshots} and \textit{graph deltas}.
\subsection{Snapshots and Deltas}
We consider an element, node or edge, of a graph $G$ as \textit{valid} for the time periods for which the corresponding item (user or friendship) of the social network it represents is also valid. Each node $v_i \in V$ is valid for the time periods for which the corresponding user $u_i$ participates in the social network represented by the graph. Similarly, each edge $(v_i,v_j) \in E$ is valid for the time periods that the corresponding users $u_i$ and $u_j$ are friends in the network.

\begin{definition}[Graph Snapshot]
A graph snapshot of a graph $G$, at a time point $t$, is defined as the graph 
$SG_t=(V',E')$, where $V'\subseteq V$ and $E'\subseteq E$, such that $v_i \in V'$, 
if and only if, $v_i$ is valid at time point $t$ and $(v_i,v_j) \in E'$, if and only if, $(v_i,v_j)$ is valid at time point $t$.  
\end{definition}

Graph $G$ captures the social network as it evolves.
Any update in the social network is directly reflected on $G$. 
A graph snapshot $SG_t$ of $G$ can be simply viewed as an instance of $G$ frozen at time point $t$,
capturing the state of $G$  at this specific time point.

We focus on the structure of the social network and thus consider the following 
four basic update operations that affect its structure: 
(1) the addition of a new user $u_i$ in the social network, 
(2) the creation of a new friendship relationship between two users $u_i$ and $u_j$ that were not friends, 
(3) the removal of an existing user $u_i$ and  
(4) the deletion of an existing friendship relationship between two users $u_i$ and $u_j$
that were friends. The corresponding operations in $G = (V, E)$ are:
\begin{enumerate}
\item $addNode(v_i)$ that adds a new node $v_i$ in $V$.
\item $addEdge(v_i,v_j)$ that creates a new edge $(v_i,v_j)$ between $v_i$ and $v_j$ in $E$.
\item $remNode(v_i)$ that deletes $v_i$ from $V$ and all edges that involve $v_i$ from $E$.
\item $remEdge(v_i,v_j)$ that deletes edge $(v_i,v_j)$ from $E$.
\end{enumerate}

Given two graph snapshots $SG_{t_k}$ and $SG_{t_l}$ of 
a graph $G$, we maintain in deltas the operations that, if applied to  
$SG_{t_k}$, produce $SG_{t_l}$. 
In the spirit of \cite{marian:change}, let us consider first, deltas as sets.

\begin{definition}[Delta]
Given two snapshots $SG_{t_k}=(V_k,E_k)$ and $SG_{t_l}=(V_l,E_l)$ of 
a graph $G$, delta $\Delta_{t_k,t_l}$ is a set of operations with the following properties:
\begin{enumerate}
\item $\forall \, v_i$, s.t.,  $v_i \notin V_k$ and $v_i \in V_l$, $addNode(v_i)$ $\in$ $\Delta_{t_k,t_l}$.  
\item $\forall \, v_i$, s.t., $v_i  \in V_k$ and $v_i \notin V_l$, $remNode(v_i)$ $\in$ $\Delta_{t_k,t_l}$.
\item $\forall \, (v_i,v_j)$, \,\, s.t., \,\, $(v_i, v_j) \,\,\, \notin \, E_k$ \,\, and \, $(v_i, v_j)\, \in \, E_l$, \\
$addEdge(v_i,v_j)$ $\in$ $\Delta_{t_k,t_l}$.
\item $\forall \, (v_i,v_j)$, s.t., $(v_i, v_j) \in E_k$, $(v_i, v_j)\notin E_l$, $v_i$ $\in$ $V_l$ and $v_j$ $\in$ $V_l$, 
$remEdge(v_i,v_j)$ $\in$ $\Delta_{t_k,t_l}$.
\end{enumerate}
These are the only operations that appear in $\Delta_{t_k,t_l}$.
\end{definition}

$\Delta_{t_k,t_l}$  is the unique minimal 
set needed for deriving $SG_{t_l}$ from $SG_{t_k}$, since it does not contain any redundant operations
and all its operations are necessary for producing $SG_{t_l}$.

\begin{lemma}
$\Delta_{t_k,t_l}$ is unique and minimal.
\end{lemma}

Applying deltas on graph snapshots is denoted using $\circ$: $\Delta_{t_k,t_l}\circ SG_{t_k}=SG_{t_l}$.
We say that a delta $\Delta_{t_k,t_l}$  is a {\em forward delta}, if $t_k<t_l$, 
i.e., when it includes operations to be applied on an older snapshot  
to create a more recent one.  

In our current approach, we record all update operations.
Thus, our deltas may contain redundant operations. 
For example, consider an edge $(v_i, v_j)$ that represents a friendship relationship  
created between $u_i$ and $u_j$ and later deleted.
We maintain both corresponding $addEdge$ and $remEdge$ operations, 
since we want to be able to retrieve all snapshots, including the one when $(u_i, u_j)$ was valid. 
We maintain such deltas as sets of operations annotated with the time point 
at which the operation occurred.
Updates are recorder as they happen in the social network, ``forward'' in time.
We call such deltas \textit{Interval Deltas}.

\begin{definition}[Interval Delta]
For a graph $G$ and 
a time interval $[t_0, t_{cur}]$, an interval delta $\Delta_{[t_0, t_{cur}]}$ is a set of pairs, $(op, t)$, 
such that a pair $(op, t)$ $\in$ $\Delta_{[t_0, t_{cur}]}$, if and only if, operation $op$ appeared 
in $G$ at time point $t$ $\in$ $[t_0, t_{cur}]$. 
\end{definition}

In the rest of this paper, 
we refer to interval deltas as deltas for simplicity as this is the only type of deltas we use in our approach.
Since we record all update operations in the time interval, our deltas are 
not minimal. 
As explained, redundant information is required for being able to retrieve a snapshot 
for any time point in the interval.
Formally, we want to ensure that our deltas  are \textit{complete}. 

\begin{definition}[Complete Delta]
A delta $\Delta_{[t_0, t_{cur}]}$ \\
is complete, if given the graph snapshot $SG_{t_0}$, 
we can derive any snapshot $SG_{t'}$, $t' \in [t_0,t_{cur}]$, 
by applying the operations $op$s of $\Delta_{[t_0,t_{cur}]}$ 
for which $t<t'$, that is if we apply $\Delta_{[t_0,t']}\subseteq\Delta_{[t_0,t_{cur}]}$:

\centerline{$\Delta_{[t_0,t']}\circ SG_{t_0}=SG_{t'}$} 
\end{definition}

For complete deltas, to reconstruct any snapshot, we just need 
an initial snapshot and the delta. 
Algorithm \ref{alg:forRec} presents the reconstruction process, assuming
for simplicity, that operations in the deltas are ordered by time.

\begin{algorithm}[t]
	\caption{$ForRec(SG_{t_0}, \Delta_{[t_0,t_{cur}]}, t')$}
	\small
	\label{alg:forRec}
	\begin{algorithmic}[1]
		\REQUIRE{$SG_{t_0}$, $\Delta_{[t_0,t_{cur}]}$, $t' \in [t_0,t_{cur}]$}
		\ENSURE{$SG_{t'}$}
		\vspace{0.1cm}
		\hrule
		\vspace{0.1cm}
		\STATE {$t:=t_0$}
		\STATE {copy $SG_{t_0}$ to $SG_{t'}$}
	   \STATE {Start from the end of $\Delta_{[t_0,t_{cur}]}$}
		\WHILE{$t<t'$}
				\STATE {Read next operation $op$ and its time $t$ in $\Delta_{[t_0,t_{cur}]}$}
				\IF {$op=addNode(v_i)$}
					\STATE{add new node $v_i$ in $SG_{t'}$}
				\ELSIF {$op=addEdge(v_i,v_j)$}
					\STATE {Find $v_i$ and $v_j$ in $SG_{t'}$}
					\STATE {add new edge $(v_i,v_j)$}
				\ELSIF{$op=remNode(v_i)$}
					\STATE {Find $v_i$ in $SG_{t'}$}
					\STATE {Remove $v_i$ from $SG_{t'}$}
				\ELSE
						\STATE {Find $v_i$ and $v_j$ in $SG_{t'}$}
					\STATE {remove edge $(v_i,v_j)$}
			\ENDIF
		\ENDWHILE
		\RETURN {$SG_{t'}$};
	\end{algorithmic}
\end{algorithm}

\noindent\textbf{Inverted Deltas.} 
So far we have considered only a forward application of deltas. 
Let us now consider the case where we want to move ``\textit{backwards}'' in time. 
That is, given a snapshot $SG_{t_k}$ at $t_k$, we want to retrieve a
snapshot $SG_{t_l}$ at $t_l$, where $t_l$ $<$ $t_k$. 
To achieve this, we define an inverted delta and apply this delta on $SG_k$.

\begin{definition}[Inverted Delta ($\bar{\Delta}$)]
Given a graph snapshot  $SG_{t_{cur}}$ and $\Delta_{[t_0,t_{cur}]}$, we define 
the inverted Delta, $\bar{\Delta}_{[t_0,t_{cur}]}$, to be the set of operations such that:  

\centerline{for each $t' \in [t_0,t_{cur}]$,  $\bar{\Delta}_{[t',t_{cur}]} \circ SG_{t_{cur}} = SG_{t'}$}
\noindent and $\bar{\Delta}_{[t',t_{cur}]}\subseteq \bar{\Delta}_{[t_0,t_{cur}]}$.
\end{definition}

To invert our deltas, we apply the reverse operation for each of the operations they include. In particular:
\begin{enumerate}
\item  $\overline{addNode(v_i)}=remNode(v_i)$.
\item $\overline{addEdge(v_i,v_j)}=remEdge(v_i,v_j)$.
\item $\overline{remNode(v_i)}=addNode(v_i)$.
\item $\overline{remEdge(v_i,v_j)}=addEdge(v_i,v_j)$.
\end{enumerate}
All operations can be inverted as long as the necessary information is maintained in the 
forward delta. 
In particular, to maintain a complete delta that 
is also {\em invertible}, we make the following assumption. 
Before recording any $remNode(v_i)$ in the delta, 
we record first $remEdge(v_i,v_j)$ operations, 
for each edge of $v_i$, annotated with the same time point 
as the  $remNode(v_i)$ operation.

Algorithm \ref{alg:backRec} presents the backward reconstruction procedure that given a graph snapshot derives a previous one by inverting the delta file.
\begin{algorithm}[t]
	\caption{$BackRec(SG_{t_{cur}}, \Delta_{[t_0,t_{cur}]}, t')$}
	\small
	\label{alg:backRec}
	\begin{algorithmic}[1]
		\REQUIRE{$SG_{t_{cur}}$, $\Delta_{[t_0,t_{cur}]}$, $t' \in [t_0,t_{cur}]$}
		\ENSURE{$SG_{t'}$}
		\vspace{0.1cm}
		\hrule
		\vspace{0.1cm}
		\STATE {$t:=t_{cur}$}
		\STATE {copy $SG_{t_{cur}}$ to $SG_{t'}$}
		\STATE {Open $\Delta_{[t_0,t_{cur}]}$ and start reading from its beginning}
		\WHILE{$t>t'$}
				\STATE {Read next operation $op$ and its time $t$ in $\Delta_{[t_0,t_{cur}]}$}
					\STATE {Apply $\overline{op}$ at $SG_{t'}$}
		\ENDWHILE
		\RETURN {$SG_{t'}$};
	\end{algorithmic}
\end{algorithm}
 
\subsection{Storage and Maintenance}
We maintain forward, complete and invertible deltas. Let us now discuss 
issues regarding the efficient reconstruction of snapshots using such deltas.

\begin{theorem}
For a graph $G$, given a delta $\Delta_{[t_0,t_{cur}]}$, 
if the delta is complete and invertible, 
to reconstruct a graph snapshot of $G$ at any time point $t \in [t_0,t_{cur}]$, 
it suffices to maintain only one graph snapshot. 

\vspace*{0.1in}
\noindent \textnormal{\textbf{Proof}. Let $SG_t$, $t \in [t_0,t_{cur}]$, be the graph snapshot we maintain. For a time point $t_k \in [t_0,t_{cur}]$ such that $t_k > t$, we reconstruct $SG_{t_k}$ with forward reconstruction by applying $\Delta_{[t,t_k]} \subseteq \Delta_{[t_0,t_{cur}]}$ on $SG_t$. For a time point  $t_l \in [t_0,t_{cur}]$ such that $t_l < t$, we reconstruct $SG_{t_l}$ with backward reconstruction by applying $\bar{\Delta}_{[t_l,t]} \subseteq \bar{\Delta}_{[t_0,t_{cur}]}$ on $SG_t$. }
\end{theorem}

Thus, based on Theorem 1, to capture the evolution of $G$ and support historical queries, 
it suffices to maintain either the original graph snapshot $SG_{t_0}$, 
since with forward reconstruction, we can derive any snapshot $SG_t$, or the current graph snapshot $SG_{t_{cur}}$, 
since with backward reconstruction, we can derive again any $SG_t$. 
The only difference between these two approaches is the cost required for reconstructing 
$SG_t$. 

If we assume that the cost of applying either the delta or the inverted delta on a snapshot is the same, 
the main factor that influences the reconstruction cost is the amount of operations (and their type) 
that we need to apply on the given snapshot. 
Therefore, it is easy to see that maintaining the original snapshot $SG_{t_0}$ 
is more appropriate when we expect more queries about the past, 
while maintaining $SG_{t_{cur}}$ is more appropriate when we expect queries about the more recent past to be more popular.

In our work, we follow the second approach, since this
approach supports queries on the current graph snapshot more efficiently. 
Thus, we maintain the current snapshot $SG_{t_{cur}}$ and $\Delta_{[t_0,t_{cur}]}$. 
As updates occur in $G$, we need to update 
both the current snapshot and the delta. 
Algorithm \ref{alg:upd} describes the update procedure. It uses an 
additional temporary delta that records the updates on $G$ until 
the next time unit and then applies this delta on the current snapshot 
to derive the next current snapshot. The algorithm is applied anew for the next time unit, and so on.

\begin{algorithm}[t]
	\caption{$Update(G, SG_{t_{cur}}, \Delta_{[t_0,t_{cur}]})$}
	\small
	\label{alg:upd}
	\begin{algorithmic}[1]
		\REQUIRE{$G, SG_{t_{cur}}$, $\Delta_{[t_0,t_{cur}]}$}
		\ENSURE{$SG_{t_{cur+1}}, \Delta_{[t_0,t_{cur+1}]} $}
		\vspace{0.1cm}
		\hrule
		\vspace{0.1cm}
		\STATE {Initialize $\Delta'_{[t_{cur},t_{cur+1}]}$}
		\WHILE {Time point $t \in [t_{cur},t_{cur+1}]$}
		\FORALL {Update operations $op$ on $G$ in $t \in [t_{cur},t_{cur+1}]$}
		\STATE {Record $(op,t)$ in $\Delta'_{[t_{cur},t_{cur+1}]}$}
		\ENDFOR
		\ENDWHILE
		\STATE {$SG_{t_{cur+1}}=\Delta'_{[t_{cur},t_{cur+1}]} \circ SG_{t_{cur}}$}
		\STATE {Append $\Delta'_{[t_{cur},t_{cur+1}]}$ at the end of $\Delta_{[t_0,t_{cur}]}$ to get $\Delta_{[t_0,t_{cur+1}]}$}
		\RETURN {$SG_{t_{cur+1}}, \Delta_{[t_0,t_{cur+1}]} $};
	\end{algorithmic}
\end{algorithm}
\noindent\textbf{Materializing Snapshots.}
While maintaining a single snapshot and the delta 
suffices for reconstructing any graph snapshot in the time interval covered by the delta, 
such reconstruction may not be efficient. 
As time progresses and more update operations occur, 
deltas grow in size and applying large parts of them to reconstruct past snapshots 
may become very costly. For instance, if we want to reconstruct the original graph snapshot,
we have to apply the entire delta file that 
may include update operations that have occurred in a social network over months or even years.

To improve efficiency,
we propose  materializing and maintaining intermediate graph snapshots
in addition to the current snapshot $SG_{t_{cur}}$.
Let $S$ be the  sequence,  $SG_{t_{i_1}}$, $\dots$ $SG_{t_{i_m}}$, $SG_{t_{cur}}$, $m$ $\geq$ 1, of 
the available materialized snapshots. To reconstruct a snapshot $SG_{t_k}$, 
we would like to start our reconstruction from the snapshot in
this sequence that would result in the most efficient reconstruction.
We consider different approaches on how to select the most appropriate snapshot, $SG_{t_{l}}$ $\in$ $S$, 
for reconstructing a snapshot $SG_{t_k}$. 

\vspace*{0.1cm}
\noindent \textit{Time-based selection.} 
Given the sequence $S$ of materialized snapshots, the snapshot $SG_{t_l}$ 
is defined as the one closest in time to $t_k$, i.e., the one 
with the smallest  ${|t_k-t_l|}$ value over all 
$t_l \in [t_0,t_{cur}]$ for which we have materialized snapshots available.  

\vspace*{0.1cm}
\noindent \textit{Operation-based selection.} 
Given the sequence $S$ of materialized snapshots, 
the snapshot $SG_{t_l}$ is defined as the one for which the operations in 
the delta ($\Delta_{[t_l,t_k]}$, if $t_l<t_k$, or  $\Delta_{[t_k,t_l]}$ if $t_l>t_k$) 
that need to be applied on $SG_{t_l}$ to derive $SG_{t_k}$ are the minimum over all the 
other deltas corresponding to the other snapshots in $S$.

\vspace*{0.1cm}
Regardless of the selection method chosen, 
depending on whether $t_l<t_k$ or $t_l>t_k$, 
we need to apply respectively forward or backward reconstruction 
using the corresponding part of the delta. 

Time-based selection can be applied more efficiently, 
as we only need to determine the snapshot closest in time to $SG_{t_k}$. 
However, if the update operations are not uniformly distributed through time, 
as is usually the case in social networks where churns of activity occur often, 
the selected snapshot $SG_{t_l}$ is not the most appropriate one.

On the other hand, operation-based selection requires that we measure 
the number of operations on the corresponding $\Delta_{[t_l,t_k]}$ if $t_l<t_k$ or  
$\Delta_{[t_k,t_l]}$ if $t_l>t_k$ which induces an additional cost. 
However, this selection guarantees that the selected snapshot yields the best 
cost for the reconstruction process as it requires the minimum number of operations to be applied.

To facilitate this process, we may also assume that deltas are 
split into disjoint intervals. In particular, along with 
each snapshot $SG_{i_j}$ in the sequence of materialized snapshots, we may maintain 
a delta $\Delta_{[t_{i_{j-1}}, t_{i_j}]}$ reporting the update
operations from the snapshot preceding it in the sequence.

\noindent \textbf{Discussion.}
An important issue that arises is when do we  
materialize a graph snapshot, i.e., how do we select the time
points at which we materialize the next snapshot 
in the sequence. 

A straightforward approach 
is to materialize snapshots \textit{periodically}, 
e.g., take one snapshot per hour, day or month. 
However, this solution has the same problem with the time-based selection of snapshots, i.e., 
it assumes that changes in a social graph occur uniformly through time.

Similarly to operation-based selection,
an alternative approach is to determine whether to materialize the next snapshot or not based 
on the amount of update operations that have occurred. 
Thus, time periods with many changes would be represented with more snapshots than
time periods with fewer changes.

Finally, snapshot materialization can be based on the similarity between snapshots. 
If two snapshots in successive time periods are similar, then we do not need to materialize both, 
whereas, if they differ significantly, then we could materialize both. 
While at first, this approach seems similar to determining the next materialized snapshot 
based on the number of update operations, they are not the same.
A snapshot may not be very different from a previous one, even if 
many operations have occurred, if such 
operations reverse themselves, e.g. the same nodes join and leave the graph repeatedly.

\begin{table*}
\centering
\caption{Examples of query types}
\label{tab:types}
\begin{tabular}{|l|l|l|l|}
\hline
\multicolumn{2}{|c|}{\backslashbox{Time}{Graph}}& Node-centric & Global\\
\hline
\multicolumn{2}{|c|}{Point}& the degree of $v_i$ at $t_k$& the diameter of $G$ at $t_k$  \\
\hline
\multirow{2}{*}{Range}& Differential & how much the degree of  $v_i$ changed in $[t_k,t_l]$& how much 
the diameter of $G$ changed in $[t_k,t_l]$\\
\cline{2-4}
& Aggregate& average degree of $v_i$ in $[t_k,t_l]$& average diameter of $G$ in $[t_k,t_l]$\\
\hline
\end{tabular}
\end{table*}

\section{Evaluating Historical Queries}
So far, we have discussed the problem
of reconstructing snapshots, given one or more materialized graph snapshots and deltas.
In this section, we address 
the problem of evaluating historical queries. 


\subsection{Query Types}
Historical graph queries can be categorized along two dimension: 
time and the part of the graph they involve. 
With regards to the \textit{time dimension}, 
queries can be further distinguished into point queries
and range queries. 
\textit{Point queries} refer to a single point in time, for example,
what is the degree of node $v_i$ at $t_k$, that may correspond
to asking for the number of friends that user $u_i$ had at this specific time in the past.
\textit{Range queries} refer to a time interval or a set of time intervals and
can be further classified as differential or aggregate.
\textit{Differential range queries} evaluate how much a measure changes during a time interval. 
An example such query is asking how much the degree of node $v_i$  
changed in $[t_k, t_l]$, that may correspond to asking about the change of popularity of
user $u_i$ in this time interval. 
Finally, \textit{aggregate range queries} evaluate an aggregate function over a time interval. 
An example such query is looking for the average degree of node $v_i$ in $[t_k, t_l]$,
that may correspond to asking for the average number of friends that user $u_i$ had in this  
interval.

With regards to the \textit{part of the graph}, we distinguished
queries as either node-centric or global queries \cite{KTSLF11}. 
{\em Node-centric queries} are queries that involve one or a few nodes of the graph.  
The degree query that we have used as an example for the time dimension is a node-centric query.
The main characteristic of such queries is that their evaluation does not 
require traversing the entire graph but only accessing a subgraph targeted 
by the query. Other node-centric queries include neighborhoods, induced subgraphs, 
and K-core queries. \textit{Global queries} are queries that 
refer to properties of the entire graph.  
Example global queries include PageRank-based queries, 
the discovery of connected components and estimating the diameter and the degree distribution. 
Table \ref{tab:types} summarizes our query classification.


\subsection{Query Processing}
Next, we present different plans for evaluating historical queries 
in $[t_0,t_{cur}]$ on a graph $G$ 
given the current graph snapshot $SG_{cur}$ and its delta, $\Delta_{[t_0,t_{cur}]}$. 

\subsubsection{Two-Phase Query Plan}
A general strategy for evaluating any historical query $q$ for any time point or range in $[t_0,t_{cur}]$
is to reconstruct the required graph snapshots 
that are determined by the query and then evaluate $q$ on them. 
Thus, the query processing plan is a two-phase plan
that involves (1) a \textit{snapshot reconstruction phase} and (2) a \textit{query processing phase}. 
During snapshot reconstruction, 
backwards reconstruction is applied on $SG_{t_{cur}}$ to acquire the snapshots 
required for evaluating $q$. 
The query processing phase takes as input the graph snapshots generated by the first phase, 
evaluates $q$ on them and combines the results if needed so as to derive the final query result. 
This is the most general plan and can be used to evaluate 
all types of  queries as indicated in Table \ref{tab:process}. 

For example for point, node-centric queries that 
ask for evaluating a measure $m$ for a node $v_i$ (e.g., $m$ may be be the 
degree of $v_i$) at time point $t_k$, the two-phase query plan is defined as follows.

\vspace*{-0.3cm}
\begin{algorithm}[H]
	\small
	\label{alg:pnc}
	\begin{algorithmic}[1]
		\REQUIRE{$SG_{t_{cur}}$, $\Delta_{[t_0,t_{cur}]}$, $t_k \in [t_0,t_{cur}]$, $v_i$}
		\ENSURE{$m(v_i)$}
		\vspace{0.1cm}
		\hrule
		\vspace{0.1cm}
		\STATE {$SG_{t_k}$=BackRec($SG_{t_{cur}}$, $\Delta_{[t_0,t_{cur}]}$, $t_k$)}
		\STATE {evaluate $m(v_i)$ on $SG_{t_k}$}
		\RETURN {$m(v_i)$};
	\end{algorithmic}
\end{algorithm}
\vspace*{-0.28cm}

Now, consider a point range query with range $[t_k,t_l] \subseteq [t_0,t_{cur}]$. 
For a point differential query (e.g., how much the degree of $v_i$ has changed in $[t_k,t_l]$), 
the query plan requires the construction of two snapshots.
Note that the second snapshot, $SG_{t_{l}}$, 
is reconstructed based on the first reconstructed snapshot $SG_{t_{l}}$ 
to avoid applying the same part of  $\Delta_{[t_0,t_{cur}]}$ twice on the current graph.

\vspace*{-0.3cm}
\begin{algorithm}[H]
	\small
	\label{alg:rdnc}
	\begin{algorithmic}[1]
		\REQUIRE{$SG_{t_{cur}}$, $\Delta_{[t_0,t_{cur}]}$, $[t_k,t_l] \subseteq [t_0,t_{cur}]$, $v_i$}
		\ENSURE{$d$}
		\vspace{0.1cm}
		\hrule
		\vspace{0.1cm}
		\STATE {$SG_{t_l}$=BackRec($SG_{t_{cur}}$, $\Delta_{[t_0,t_{cur}]}$, $t_k$)}
		\STATE {$SG_{t_k}$=BackRec($SG_{t_{l}}$, $\Delta_{[t_0,t_{cur}]}$, $t_l$)}
		\STATE {evaluate $m_k(v_i)$ on $SG_{t_k}$}
		\STATE {evaluate $m_l(v_i)$ on $SG_{t_l}$}
		\STATE {$d=|m_k(v_i)-m_l(v_i)|$}
		\RETURN {$d$};
	\end{algorithmic}
\end{algorithm}
\vspace*{-0.25cm}

Let us now consider an aggregate range node-centric query (e.g., 
the average degree of $v_i$ in $[t_k,t_l]$) denoted by $F(m(v_i))$. 
This query requires the construction of a snapshot for each time unit 
in the time interval so as to compute the average between all values of $m(v_i)$ in this time range.

\vspace*{-0.3cm}
\begin{algorithm}[H]
	\small
	\label{alg:ranc}
	\begin{algorithmic}[1]
		\REQUIRE{$SG_{t_{cur}}$, $\Delta_{[t_0,t_{cur}]}$, $[t_k,t_l] \subseteq [t_0,t_{cur}]$, $v_i$}
		\ENSURE{$F(m(v_i))$}
		\vspace{0.1cm}
		\hrule
		\vspace{0.1cm}
		\FORALL{$t \in [t_k,t_l]$}
		\STATE {$SG_{t}$=BackRec($SG_{t_{cur}}$, $\Delta_{[t_0,t_{cur}]}$, $t$)}
		\STATE {evaluate $m_t(v_i)$ on $SG_{t}$}
		\ENDFOR
		\STATE {apply aggregation function $F$ on all $m_t(v_i)$}
		\RETURN {$F(m(v_i))$};
	\end{algorithmic}
\end{algorithm}
\vspace*{-0.25cm}

Similar algorithms can be used for global queries. 

For simplicity, we have assumed that reconstruction uses only the current graph snapshot.
If materialized snapshots are maintained, the only difference is that 
a selection phase is applied before the reconstruction phase.
During the {\em selection phase}, we determine the most appropriate snapshot to
be used for reconstruction and based on this selection
whether to use forward or backward reconstruction. 
If more than one snapshot need to be reconstructed for query processing, 
then a different selection and reconstruction procedure may be used for each one of them. 
An interesting problem in this case is re-using reconstructed snapshots.
A simple example was shown for point range queries. 

The two-phase query plan 
can be used to evaluate all types of historical queries. 
However, the reconstruction of snapshots can be costly. 
Next, we consider alternative plans for specific query types that avoid this phase.

\subsubsection{Delta-Only Query Plan}
With {\em delta-only query plans}, a query is evaluated directly on  
the deltas. No snapshot reconstruction is required. Furthermore, there is no need to access 
any of the snapshots. 
Such plans are applicable to differential range node-centric queries (Table \ref{tab:process}). 
For such queries one can compute how much a measure 
has changed by accessing the corresponding update operations 
in the delta file for the given time interval. 

For instance, consider a range differential node-centric query
asking  for the difference in the degree of node $v_i$ in $[t_k,t_l]$. 
This query can be evaluated with a delta-only plan, 
if one  just counts the add and remove edge operations that 
involve $v_i$ in $\Delta_{[t_k,t_l]}$. 


\subsubsection{Hybrid Query Plan}
Finally, we consider hybrid plans. 
Such plans access both the current snapshot  and the delta,
but do not require the reconstruction of any graph snapshots. 

These plans are applicable to point and aggregate range node-centric queries (Table \ref{tab:process}).
For instance, consider an aggregate range node-centric query, e.g., 
asking for the average degree of $v_i$ in $[t_k,t_l]$. 
The hybrid plan evaluates the degree of $v_i$ on $SG_{cur}$ and then 
traverses $\Delta_{[t_k,t_l]}$ to compute the degree at each time unit in the 
requested range. Then, its average is computed.

Note, that there are cases in which more than one pass of the $\Delta_{[t_k,t_l]}$ 
may be required to evaluate a node-centric measure. 
For instance, consider a query for the average degree of the induced subgraph of $v_i$ in $[t_k,t_l]$, 
where the induced subgraph of $v_i$ is the subgraph formed by $v_i$ and its neighbors. 
By traversing the delta, we may add new nodes in the subgraph and therefore  
need to go back and include edges of these specific new nodes 
that have not been included initially, as they were not part of the original subgraph. 

\begin{table}
\centering
\caption{Query processing}
\label{tab:process}
\begin{tabular}{|l|l|l|l|l|}
\hline
\multicolumn{2}{|c|}{Query Types} &\multicolumn{3}{|c|}{Query Plans}\\
\hline
& & Two& Delta & Hybrid\\
& & Phase & only & \\
\hline
\multirow{2}{*}{Point} &Node-centric& \checkmark & & \checkmark \\
 \cline{2-5}
&Global&\checkmark & & \\
\hline
Range &Node-centric& \checkmark &\checkmark &\checkmark\\
 \cline{2-5}
differential &Global&\checkmark & &\\
\hline
Range &Node-centric& \checkmark && \checkmark\\
 \cline{2-5}
aggregate &Global&\checkmark &&\\
\hline
\end{tabular}
\end{table}

\subsection{Delta Indexing and Optimizations}
As pointed out, snapshot reconstruction is the most costly phase in our query plans 
and we would rather avoid it whenever possible. 
For queries for which reconstruction cannot be avoided, we present techniques that improve its efficiency.

\subsubsection{Partial Reconstruction}
The difference between node-centric and global queries is that  
node-centric queries do not require traversing the entire 
graph but are targeted to one or a few nodes. 
Let $G'=(V',E')$ be the sub-graph of $G$ that a node-centric query $q$ 
needs to access. 
Then, instead of reconstructing the snapshots $SG_t$ that $q$ requires of the entire graph $G$, 
it suffices to reconstruct the corresponding snapshots of the subgraph $G'$. 
During snapshot reconstruction, all add and remove operations involving nodes and edges 
such that $v_i \notin V'$ and $(v_i,v_j) \notin E'$ are ignored. 
Multiple passes of the delta may be required to determine all elements to be included in 
the snapshots.

\subsubsection{Indexing}
To further improve efficiency during reconstruction, we propose 
building indexes on the delta. 
Indexing also improves delta-based query plans 
by enabling faster access to specific parts of the delta. 
Indexing may  improve performance significantly, 
especially considering that the size of a delta grows constantly though time.

\noindent\textbf{Temporal Index.}
Snapshot reconstruction and query processing 
usually require applying or accessing a part $\Delta_{[t_k,t_l]} \subseteq \Delta_{[t_0,t_{cur}]}$ 
of the delta file to reconstruct graph snapshots or evaluate a query. 
Therefore, using a \textit{temporal index} improves the efficiency of 
these procedures by enabling faster access to the desired parts of the delta. 

\noindent\textbf{Node-centric Index.}
Besides temporal indexing, another option is to apply \textit{node-centric} indexing 
to enable the efficient location of all operations  
associated with a specific node. A node-centric index improves the evaluation 
of node-centric queries for all three query plans that we have discussed. 
It also facilitates partial reconstruction. 

\section{Preliminary Evaluation}
The efficiency of processing historical queries depends on the underlying
storage model.
In our initial implementation, deltas are stored
in append-only files. Any materialized graph snapshots are stored in a native graph database.
We also maintain an in-memory node-centric index on the delta file.
The use of a native graph database results in faster execution of graph queries,
but it does not support any form of locality. 

The goal of this preliminary evaluation is to present some initial quantitative results regarding the 
efficiency of the proposed query plans.
To this end, we run a node-centric query that asks for the degree of a 
random node $v$ at time point $t$.
We evaluate the following four different plans: (a) a two-phase query plan without indexing (two-phase), 
(b) a hybrid query plan without indexing (hybrid), (c) a two-phase approach with indexing (two-phase-index) 
and (d) a hybrid approach with indexing (hybrid-index). 
For the two-phase approach, we used partial reconstruction.
We also used only the current snapshot and backward reconstruction (no additional materialized
snapshots were used). 

We generate graphs that are scale-free, in an effort to mimic the form of online social network graphs. 
To generate scale-free graph snapshots, we use the method in \cite{ren:evolving} 
that extends the Barabasi algorithm \cite{barabasi} for generating successive scale-free graphs. 
Table~\ref{table1} summarizes the characteristics of the synthetic dataset.
%
All algorithms are implemented in Java. 
The experiments were run on a Linux Machine with 2.8GHz Dual Core Intel and 4GB of memory.
As our native graph database, we used Neo4j \footnote{http://neo4j.org}.

\begin{figure}
	\centering
		\includegraphics[width=4.5cm]{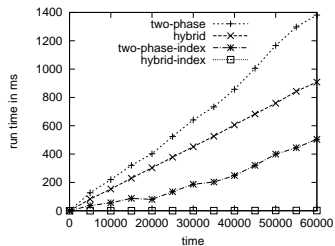}
		\vspace{-0.1cm}
	\caption{Run time in ms for executing a degree query at different time points (time measured in operations).} 
	\vspace{-0.3cm}
	\label{figure1}
\end{figure}

Figure ~\ref{figure1} reports the run time  
in milliseconds for executing the query at different time points. The time  in the x-axis (measured 
in number of operations) proceeds backwards (i.e., point 0 corresponds to the current snapshot). 
The more time passes from the current snapshot, the more 
expensive is to evaluate the query, since reconstructing the past
snapshot requires the application of more operations.
The two-phase algorithm takes the most amount of time, 
due to the cost of the reconstruction phase. 
This phase is especially expensive, since in Neo4j, any modification to stored
data is associated with a transaction and is flashed directly to disk. 
The usage of a node-centric index on the delta file 
leads to significant gains for both the two-phase and the hybrid approach.

\begin{table}
\centering
\small
\caption{Synthetic dataset}
  \begin{tabular}{ | c | c |}
    \hline
   number of inserted nodes & 5063  \\ \hline
   number of inserted edges &  41067 \\ \hline
   number of removed edges & 18280 \\ \hline
   size of delta file	& 64410 operations/1.3 MB \\
    \hline
  \end{tabular}
  \label{table1}
\end{table}

\section{Related Work}
There is a large body of work on temporal data management including relational databases (see, 
for example \cite{OS95} and \cite{ST99} for excellent surveys on the topic), 
RDF (e.g., \cite{gutierrez:rdf}) and XML documents (e.g., \cite{marian:change}, \cite{BKTT04}). 
Although maintaining deltas has also been used in such cases, 
the large scale and the logical model, being in our case a graph, introduces new problems. 

Collecting a sequence of versions of XML documents 
from the web is considered in \cite{marian:change}. 
The difference between two consecutive versions is 
computed and represented by complete deltas based on persistent identifiers 
assigned to each XML node, 
while only the current version of the document is maintained. 
To avoid the overhead of applying deltas to retrieve previous versions, in \cite{BKTT04}, 
they merge all versions of XML data into one hierarchy where an element appearing in multiple versions is stored only once along with a timestamp. 
To handle temporal RDF data, temporal reasoning is incorporated into RDF in \cite{gutierrez:rdf}, thus yielding temporal RDF graphs. Semantics were also defined for these graphs which include the notion of temporal entailment as well 
as a syntax to incorporate this framework into standard RDF graphs by adding temporal labels. 
Clearly, our approach is different in that it considers time with respect to graph evolution. 

Numerous algorithms and data structures have been proposed for processing graph queries on large graphs. GBASE \cite{KTSLF11} and Pregel \cite{MABDHLC10} are two 
general graph management systems that work in parallel and 
distributed settings and support large-scale graphs for various applications. GBASE is based on a common underlying primitive of several graph mining operations, which is shown to be a generalized form of matrix-vector multiplication \cite{KTF11}, while Pregel  is based on a sequence of supersteps that are applied in parallel by each node executing the same user-defined function that expresses the logic of a given algorithm and are separated by global synchronization points.
In future work, we plan to explore such techniques for reconstructing snapshots in parallel.

The most relevant to our work is perhaps the historical graph structure recently proposed 
in \cite{ren:evolving}. The authors consider a sequence of graphs produced as the graph evolves over time. Since the graphs in the sequence are very similar to each other, they propose computing graph representatives by clustering similar graphs and then storing appropriate differences from these representatives, 
instead of storing all graphs. 
Our approach is different in that we want to support a broad range of historical queries, not just queries that involve a single snapshot graph.

There is also a large body of work that studies the evolution of real-world networks over time. 
In \cite{LKF05}, it was shown that for a variety of real-world networks,  graph density increases 
following a power-low and the graph diameter shrinks. 
Works on specific networks such as Flickr \cite{grwosn} and Facebook \cite{acwosn} 
also study network growth and their results can be exploited to enrich our model.


\section{Conclusions}
In this paper, we presented a model 
for capturing 
graph evolution through time  based on the use of graph snapshots and deltas. 
We showed how by maintaining only the current graph snapshot and a delta, 
we can reconstruct any past graph snapshot. 
Then, we introduced a general two-phase query plan 
based on snapshot reconstruction to evaluate any historical query as well as a couple of 
more efficient plans that avoid reconstruction for specific queries. 
Finally, we presented preliminary experimental results. 

\bibliographystyle{abbrv}

\small
\bibliography{sigproc} 
%
%
\end{document}